\documentclass{aa}
\usepackage{graphicx}

\usepackage{amsmath}

\def\mean#1{\left< #1 \right>}

\begin{document}

\title{Pitfalls of statistics-limited X-ray polarization analysis}
\author{V. Mikhalev}
\institute{KTH Royal Institute of Technology, Department of Physics, 106 91 Stockholm, Sweden.\\
The Oskar Klein Centre for Cosmoparticle Physics, AlbaNova University Centre, 106 91, Stockholm, Sweden.\\
\email{mikhalev@kth.se}}
\date{Accepted 23/03/2018}

\abstract
{One of the difficulties with performing polarization analysis is that the mean polarization fraction of sub-divided data sets is larger than the polarization fraction for the integrated measurement. The resulting bias is one of the properties of the generating distribution discussed in this work. The limitations of Gaussian approximations in standard analysis based on Stokes parameters for estimating polarization parameters and their uncertainties are explored by comparing with a Bayesian analysis. The effect of uncertainty on the modulation factor is also shown, since it can have a large impact on the performance of gamma-ray burst polarimeters. Results are related to the minimum detectable polarization (MDP), a common figure of merit, making them easily applicable to any X-ray polarimeter.}
{The aim of this work is to quantify the systematic errors induced on polarization parameters and their uncertainties when using Gaussian approximations and to show when such effects are non-negligible.}
{The probability density function is used to deduce the properties of reconstructed polarization parameters. The reconstructed polarization parameters are used as sufficient statistics for finding a simple form of the likelihood. Bayes theorem is used to derive the posterior and to include nuisance parameters.}
{The systematic errors originating from Gaussian approximations as a function of instrument sensitivity are quantified here. Different signal-to-background scenarios are considered making the analysis relevant for a large variety of observations. Additionally, the change of posterior shape and instrument performance (MDP) due to uncertainties on the polarimeteric response of the instrument is shown.}
{}

\maketitle

\section{Introduction}

The first observations of astrophysical X-ray polarization were made more than forty years ago (Novick et al.~\cite{Novick1972}; Weisskopf et al.~\cite{Weisskopf1976}). The field has been reinvigorated in the past decade by a series of measurements by satellite, including INTEGRAL/SPI (Dean et al.~\cite{Dean2008}; Chauvin et al.~\cite{Chauvin2013SPI}), INTEGRAL/IBIS (Forot et al.~\cite{Forot2008}; Moran et al.~\cite{Moran2016}), AstroSat/CZTI (Vadawale et al.~\cite{Vadawale2018}) and IKAROS/GAP (Yonetoku et al. ~\cite{Yonetoku2011}) as well as balloon-borne instruments (PoGOLite (Chauvin et al.~\cite{Chauvin2016PoGOLite}) and PoGO+ (Chauvin et al.~\cite{Chauvin2017PoGO+})). Results from several on-going missions are expected in the near future (POLAR (Produit et al.~\cite{Produit2005}), X-Calibur (Beilicke et al.~\cite{Beilicke2014})) and a dedicated satellite mission is in development (IXPE (Weisskopf et al.~\cite{Weisskopf2016})). Although some instruments, for example, IXPE, will be able to make polarization measurements of astrophysical sources with high precision, where the statistical analysis becomes relatively trivial, it is always desirable to observe weaker sources or to sub-divide the data. Fine splitting of data may be necessary for understanding physical phenomena, for example, gamma-ray bursts (GRBs) and pulsars require temporal binning, nebulae require spatial binning, and spectral binning is interesting for all objects. It is therefore important to know when the number of photons is sufficient for making an accurate analysis using simple methods and when a more rigorous approach is necessary.

The parameters describing linear polarization are the polarization fraction and the polarization angle. In the frequentist approach, the major challenge in estimating the polarization fraction is that it is a positive definite quantity, meaning that a non-zero fraction is measured even for an unpolarized source, thus introducing a bias. Ways of correcting for this bias have been studied by Simmons \& Stewart (\cite{Simmons1985}) and more recently by Maier et al. (\cite{Maier2014}). A Bayesian approach was first introduced by Vaillancourt (\cite{Vaillancourt2006}) and extensively expanded upon by Quinn (\cite{Quinn2012}), where the shapes of the resulting parameter distributions are described. These preceding works focus on optical measurements of polarization for which some formulae differ from the X-ray counterpart (Kislat et al.~\cite{Kislat2015}).

This work quantifies, as a function of measurement sensitivity, the error incurred when using the conventional Stokes parameter analysis. This is shown for both the polarization fraction and angle as well as for their uncertainties. Henceforth the expression ``low statistics'' is used to indicate poor data quality having a low $S/\sqrt{N}$, where $S$ is the number of signal photons and $N$ is the total number of photons (signal and background). Since a point-source polarimeter working in the low-statistics regime is likely to have a low signal-to-background ratio $R$, whereas a GRB polarimeter is expected to have a higher $R$ (especially if the GRB is bright and short in duration), the study is conducted for different values of $R$ where meaningful. Additionally, the effects of uncertainty on the modulation factor $\mu_0$, defined as the polarization fraction measured for a 100\% polarized beam, on the performance of the polarimeter are studied. The uncertainty is typically large for GRB polarimeters which are not optimized for localization of GRBs since $\mu_0$ varies with the photon incidence angle.

\section{Parameter estimation}

For a Compton scattering (Lei et al.~\cite{Lei1997}) or photo-electric polarimeters (Bellazzini \& Muleri~\cite{Bellazzini2010}) with a uniform response, the conditional distribution for a measured polarization angle $\psi_i$ given a polarization fraction $p_0$, a polarization angle $\psi_0$ and a modulation factor $\mu_0$ follows

\begin{equation}
\begin{split}
f(\psi_i|p_0,\psi_0,\mu_0)=\frac{1}{2\pi(S+B)}\\
\times(S\times(1+\mu_0 p_0\cos{\left(2(\psi-\psi_0\right)}))+B)
\end{split}
\label{eq:modulation_curve}
,\end{equation}
where $\psi=\phi-90^\circ$ for the Compton process and $\psi=\phi$ for the photo-electric effect, $\phi$ is the measured scattering angle, $S$ is the number of signal photons and $B$ is the number of background photons. Treating signal and background as latent variables is beyond the scope of this paper so it is assumed that the background is unpolarized and is known with much higher precision than the polarization parameters. The modulation factor $\mu_0$ is a property of the polarimeter and is derived from calibration. In what follows, the notation with subscript "$0$" means a physical parameter generating the data, in this case a set of measured scattering angles transformed to the polarization frame $\{\psi_i\}$, while subscript "r" denotes reconstructed parameters from this set. The reconstructed parameters are actually sufficient statistics for the data set allowing it to be represented by 2 scalars rather than a set of $N$ angles.

The two most common ways of computing $p_{\mathrm{r}}$ and $\psi_{\mathrm{r}}$ are by performing a $\chi^2$-fit of Eq.~\ref{eq:modulation_curve} to a histogram of scattering angles or by computing the Stokes parameters. This work only considers the latter as it avoids complicating the analytical form of the likelihood due to binning effects.

\subsection{Stokes parameters}

Polarimeters operating in the radio, infra-red, or optical domain measure photon intensities rather than individual photons as opposed to X-ray polarimeters. Clarke et al.~(\cite{Clarke1983}) discuss how the Stokes parameter distributions vary depending on the instrumental technique used for measuring these intensities. Since Eq.~\ref{eq:modulation_curve} is not used in the low-energy domain, not all results presented in that publication can be extrapolated to the X-ray energy band. In particular, the standard deviations and the correlation coefficient of Stokes parameters are affected.

In the X-ray energy band, the Stokes parameters are derived from individual photon events comprising two quantities

\begin{equation}
\begin{split}
q_i=\cos({2\psi_i})\\
u_i=\sin({2\psi_i})
\end{split}
.\end{equation}
For a total number of photons $N=S+B$ the normalized Stokes parameters are written as
\begin{equation}
\begin{split}
Q_\mathrm{r}=\frac{1}{S}\sum\limits_{i=1}^Nq_i\\
U_\mathrm{r}=\frac{1}{S}\sum\limits_{i=1}^Nu_i
\end{split}
.\end{equation}
The normalization is only proportional to the signal because the background is assumed to be unpolarized. Therefore the background contributes only to the variance of $(Q_\mathrm{r},U_\mathrm{r})$.

The Central Limit Theorem (CLT) makes $Q_\mathrm{r}$ and $U_\mathrm{r}$ Gaussian distributed as long as $N$ is sufficiently large. This is referred to here as the CLT approximation. Thus $Q_\mathrm{r}$ and $U_\mathrm{r}$ follow the bivariate Gaussian distribution
\begin{equation}
\begin{split}
B(Q_\mathrm{r},U_\mathrm{r}|Q_0,U_0)=\frac{1}{2\pi\sigma_Q\sigma_U\sqrt{1-\rho^2}}\\
\times\exp\Bigg[-\frac{1}{2(1-\rho^2)}\Bigg(\frac{(Q_\mathrm{r}-Q_0)^2}{\sigma_Q^2}+\frac{(U_\mathrm{r}-U_0)^2}{\sigma_U^2}\\
-\frac{2\rho(Q_\mathrm{r}-Q_0)(U_\mathrm{r}-U_0)}{\sigma_Q\sigma_U}\Bigg)\Bigg]
\end{split}
\label{eq:bivariate_full}
.\end{equation}
Since the mean, the standard deviation, and the correlation coefficient are sufficient statistics for Gaussian distributed data, so are $Q_\mathrm{r}$ and $U_\mathrm{r}$ together with the second moments of the data derived in Appendix~\ref{appendix_stokes_uncertainties}. Although not written explicitly in the conditioning, $Q_0$, $U_0$, $\sigma_U$, $\sigma_Q$ and $\rho$ are functions of $p_0$ and $\psi_0$.

\subsection{Polar coordinates}
Stokes parameters $(Q,U)$ are in Cartesian coordinates but can be transformed to polar coordinates to allow the polarization parameters to be expressed in terms of $(p,\psi)$. The sufficient statistics become
\begin{equation}
p_{\mathrm{r}}=2\sqrt{Q_\mathrm{r}^2+U_\mathrm{r}^2}/\mu_{\mathrm{r}}
\label{eq:define_p_r}
,\end{equation}
\begin{equation}
\psi_{\mathrm{r}}=\frac{1}{2}\arctan{\frac{U_\mathrm{r}}{Q_\mathrm{r}}}
\label{eq:define_psi_r}
,\end{equation}
as per the derivations in Appendix~\ref{appendix_stokes_parameters}.

Now Eq.~\ref{eq:bivariate_full} can be transformed to polar coordinates by using the general coordinate transformation for a probability density function (p.d.f.) yielding the likelihood
\begin{equation}
L(p_{\mathrm{r}},\psi_{\mathrm{r}}|p_0,\psi_0)=B(Q_\mathrm{r},U_\mathrm{r}|Q_0,U_0)\times|\det (J)|
\label{eq:likelihood_polar}
,\end{equation}
where $\det (J)$ is the determinant of the Jacobian given by
\begin{equation}
\begin{vmatrix}
\vspace{1mm}\frac{\partial Q_\mathrm{r}}{\partial p_\mathrm{r}} & \frac{\partial Q_\mathrm{r}}{\partial \psi_\mathrm{r}} \\ \frac{\partial U_\mathrm{r}}{\partial p_\mathrm{r}} & \frac{\partial U_\mathrm{r}}{\partial \psi_\mathrm{r}}
\end{vmatrix}=\begin{vmatrix}
\vspace{1mm}\frac{\mu_\mathrm{r}}{2}\cos(2\psi_\mathrm{r}) & -\mu_\mathrm{r}p_\mathrm{r}\sin(2\psi_\mathrm{r}) \\ \frac{\mu_\mathrm{r}}{2}\sin(2\psi_\mathrm{r}) & \mu_\mathrm{r}p_\mathrm{r}\cos(2\psi_\mathrm{r})
\end{vmatrix}=\frac{p_{\mathrm{r}}\mu_\mathrm{r}^2}{2}
.\end{equation}

Although the polarization parameters have a very similar form to that of the sufficient statistics (simply replacing the subscript "$\mathrm{r}$" by "$0$")
\begin{equation}
p_0=2\sqrt{Q_0^2+U_0^2}/\mu_0,
\end{equation}
\begin{equation}
\psi_0=\frac{1}{2}\arctan{\frac{U_0}{Q_0}}
,\end{equation}
$p_\mathrm{r}$ does not correspond to the most probable estimate of $p_0$. This occurs because $(\sigma_Q,\sigma_U,\rho)$ depend on $(p_0,\psi_0)$ and the fact that $0\leq p_0 \leq 1$ (since a polarization greater than 100\% is unphysical) but there is no upper limit on $p_{\mathrm{r}}$, that is, $0\leq p_{\mathrm{r}}$. Therefore, using the sufficient statistic $p_\mathrm{r}$ as the estimator of the polarization fraction $p_0$ incurs an error.

The error is non-neglible for the one-dimensional likelihood of $p_0$ (obtained after marginalizing over $\psi_0$) if the statistical significance of the measurement is low. It results in $\mathrm{argmax}_{p_0} L(p_\mathrm{r},\psi_\mathrm{r}|p_0)\neq p_\mathrm{r}$ and $\mean{p_\mathrm{r}}\neq p_0$, where $\mean{p_\mathrm{r}}$ is the expected value of $p_\mathrm{r}$. Hence $p_{\mathrm{r}}$, is neither the maximum likelihood nor an unbiased estimator of $p_0$. This is the case even if $\mu_\mathrm{r}=\mu_\mathrm{0}$ which occurs when there is no uncertainty on $\mu$ as is assumed throughout this section.

As the statistical precision of a measurement improves $\lim_{S \to \infty}\mean{p_{\mathrm{r}}}=p_0$ and $\lim_{S \to \infty} \mathrm{argmax}_{p_0} L(p_\mathrm{r},\psi_\mathrm{r}|p_0)= p_\mathrm{r}$. Conversely, due to symmetry, there is no bias on $\psi$ so $\mean{\psi_{\mathrm{r}}}=\psi_0$ and $\mathrm{argmax}_{\psi_0} L(p_\mathrm{r},\psi_\mathrm{r}|\psi_0)= \psi_\mathrm{r}$.

\subsection{Central Limit Theorem approximation of the likelihood}

The CLT approximation in $B(Q_\mathrm{r},U_\mathrm{r}|Q_0,U_0)$ provides an always easily computable analytical form of the likelihood given by Eq.~\ref{eq:likelihood_polar}. However, if the polarization fraction is high and the number of photons is low, the likelihood is not well described by such an approximation. The full form of the likelihood is given by a product over Eq.~\ref{eq:modulation_curve}.
\begin{equation}
\begin{split}
L(\{\psi_i\}|p_0,\psi_0,\mu_0)=\prod_{i=1}^N\frac{1}{2\pi(S+B)}\\
\times(S\times(1+\mu_0 p_0\cos{\left(2(\psi_i-\psi_0\right)}))+B)
\end{split}
\label{eq:full_likelihood}
.\end{equation}
It has $N$ factors and is therefore cumbersome to evaluate when the number of photons is large. This is unlike optical polarimetery where photon intensities of $(Q,U)$ are directly measured instead of individual scattering angles.

\begin{figure}
\resizebox{\hsize}{!}{\includegraphics{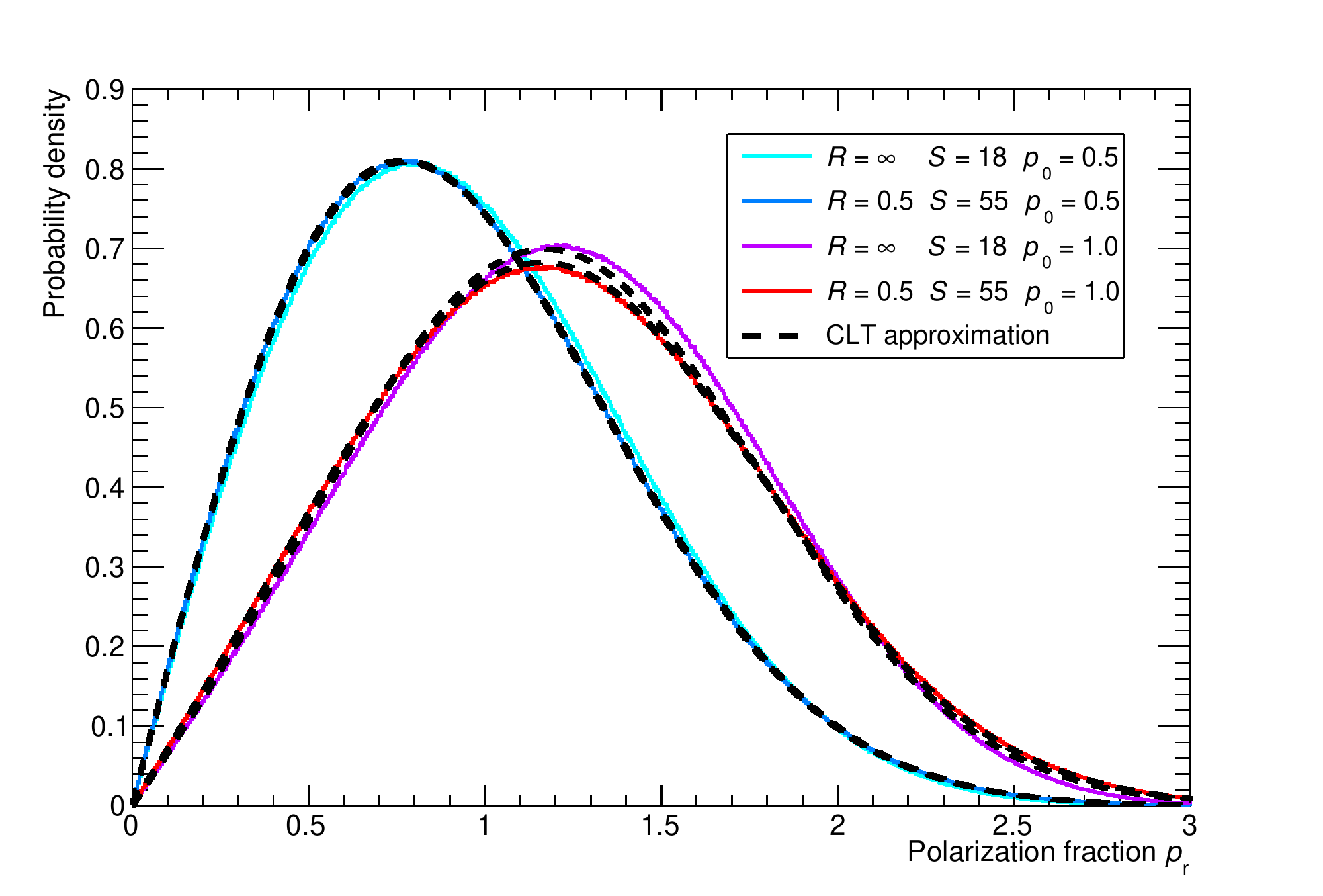}}
\caption{The comparison of the CLT approximation to the full likelihood, as given by Eqs.~\ref{eq:likelihood_polar} and \ref{eq:full_likelihood}, respectively. For $p_0=0.5,$ both pure signal and mixed scenarios are well described by the CLT approximation. For $p_0=1.0,$ the pure signal scenario deviates farther from the approximation than the scenario with background because the pure signal scenario has fewer photons. All scenarios use $\mu_0=0.5$ and $\mathrm{MDP}=2$.}
\label{fig:p_r_density}
\end{figure}

Figure~\ref{fig:p_r_density} shows the difference between the CLT approximation and $L(p_\mathrm{r}|p_0,\psi_0,\mu_0)$ as calculated by generating a data set $\{\psi_i\}$ from Eq.~\ref{eq:full_likelihood}, computing $p_\mathrm{r}$ using the Stokes parameters, repeating for many iterations and making a normalized histogram of $p_\mathrm{r}$. 

In what follows, the concept of minimum detectable polarization (MDP) at 99\% confidence level is important. The MDP (Weisskopf et al. \cite{Weisskopf2010}) is given by 
\begin{equation}
\textrm{MDP}=\frac{4.29}{\mu_0 S}\sqrt{S+B}
\label{eq:MDP}
.\end{equation}
Its statistical meaning is that, given an unpolarized source ($p_0=0$), the probability of measuring $p_{\mathrm{r}}>\textrm{MDP}$ is 1\%. This quantity is a standard figure of merit for polarimeter performance and can easily be calculated for any measurement. The number of signal photons in Fig.~\ref{fig:p_r_density} is chosen such that $\mathrm{MDP}=2$, since this is the highest ("worst") MDP considered in later sections. As seen from Fig.~\ref{fig:p_r_density}, the CLT approximation becomes more accurate as the number of photons increases and the polarization fraction $p_0$ decreases. In particular, only measurements with high signal-to-background ratio and high polarization fraction require the full likelihood given by Eq.~\ref{eq:full_likelihood}. Although the error incurred using the CLT approximation is negligible in most cases, it becomes larger when using Eq.~\ref{eq:likelihood_polar} to derive the posterior, as is discussed below. Figure \ref{fig:p_r_density} is for qualitative purposes only; the equivalent figure for $\psi_\mathrm{r}$ has been omitted since it leads to the same conclusions.

\subsection{Magnitude and importance of bias}

Bias is a frequentist concept which relies on fixing $(p_0,\psi_0)$ and investigating $\mean{p_\mathrm{r}}$. This approach provides an intuitive understanding for how an unpolarized source can produce a polarized signal ($p_\mathrm{r}>0$). Several previous measurements use the sufficient statistic $p_\mathrm{r}$ as an estimate of the polarization fraction $p_0$ (e.g., Weisskopf et al.~\cite{Weisskopf1978}; Slowikowska et al.~\cite{Slowikowska2009}) and it is therefore necessary to understand when the bias is negligible and when a more sophisticated approach is necessary.

In the frequentist approach, the p.d.f. $B(Q_\mathrm{r},U_\mathrm{r}|Q_0,U_0)$ is not a function of $\psi_0$ (it is a fixed parameter) but of $\psi_r$. It is assumed, without loss of generality due to the angular symmetry of the problem, that $\psi_0=0$ so that $\rho=0$. This can be thought of as rotating the angular coordinate system which does not have any special reference point. The definition of the "zero angle" of such a system has no influence on the polarization fraction. The p.d.f. now simplifies to the product of two normal distributions
\begin{equation}
\begin{split}
B(Q_\mathrm{r},U_\mathrm{r}|Q_0,U_0)_{\psi_0=0}=\frac{1}{2\pi\sigma_Q\sigma_U}\\
\times\exp\Bigg[-\frac{1}{2}\Bigg(\frac{(Q-\mean{Q})^2}{\sigma_Q^2}+\frac{(U-\mean{U})^2}{\sigma_U^2}\Bigg)\Bigg]
\end{split}
\label{eq:bivariate}
.\end{equation}
It can now be transformed to polar coordinates similarly to Eq.~\ref{eq:likelihood_polar} yielding
\begin{equation}
f(p_{\mathrm{r}},\psi_{\mathrm{r}}|p_0)_{\psi_0=0}=B(Q,U)_{\psi_0=0}|\det (J)|
\label{eq:likelihood_polar_fixed}
.\end{equation}
The relative mean bias $\beta$ is now given by
\begin{equation}
\beta=\frac{\mean{p_{\mathrm{r}}}-p_0}{p_0}=\frac{\int_{0}^\infty \int_{0}^{\pi} p_{\mathrm{r}} f(p_{\mathrm{r}},\Delta\psi|p_0)_{\psi_0=0}\mathrm{d}\Delta\psi\mathrm{d}p_{\mathrm{r}}-p_0}{p_0}
\label{eq:beta_nummeric}
,\end{equation}
where $\Delta\psi=\psi_{\mathrm{r}}-\psi_0=\psi_\mathrm{r}$. 

To understand which parameters have a significant impact on $\beta$ an approximate analytical expression can be derived by introducing
\begin{equation}
\sigma\equiv\frac{\sqrt{2N}}{\mu_0 S}\approx\frac{2\sigma_Q}{\mu_0}\approx\frac{2\sigma_U}{\mu_0}
.\end{equation}
It is now possible to write Eq.~\ref{eq:likelihood_polar_fixed} as
\begin{equation}
\begin{split}
f(p_{\mathrm{r}},\Delta\psi|p_0,0)=\frac{p_{\mathrm{r}}}{\pi\sigma^2} \\
\times\exp{\left(-\frac{p_{\mathrm{r}}^2+p_0^2-2p_{\mathrm{r}}p_0\cos(2\Delta\psi)}{2\sigma^2}\right)}
\end{split}
\label{eq:likelihood_polar_approx}
.\end{equation}
Integrating over $\Delta\psi$ results in
\begin{equation}
f(p_{\mathrm{r}}|p_0)=\frac{p_{\mathrm{r}}}{\sigma^2}\exp{\left(-\frac{p_{\mathrm{r}}^2+p_0^2}{2\sigma^2}\right)}\times I_0\left(\frac{p_{\mathrm{r}}p_0}{\sigma^2}\right)
\label{eq:approx_bivariate}
,\end{equation}
which is the Rice distribution where $I_0$ is the modified Bessel function of zeroth order. In the limit of high statistics, the relative mean bias is given by
\begin{equation}
\lim_{p_0/\sigma \to \infty} \beta\approx\frac{\sigma^2}{2p_0^2}=\frac{N}{S^2\mu_0^2p_0^2}=\left(\frac{\textrm{MDP}}{4.29p_0}\right)^2
\label{eq:mean_bias_mdp_approx_p0}
,\end{equation}
as shown in Appendix~\ref{appendix_relative_mean_bias}.

\begin{figure}
\resizebox{\hsize}{!}{\includegraphics{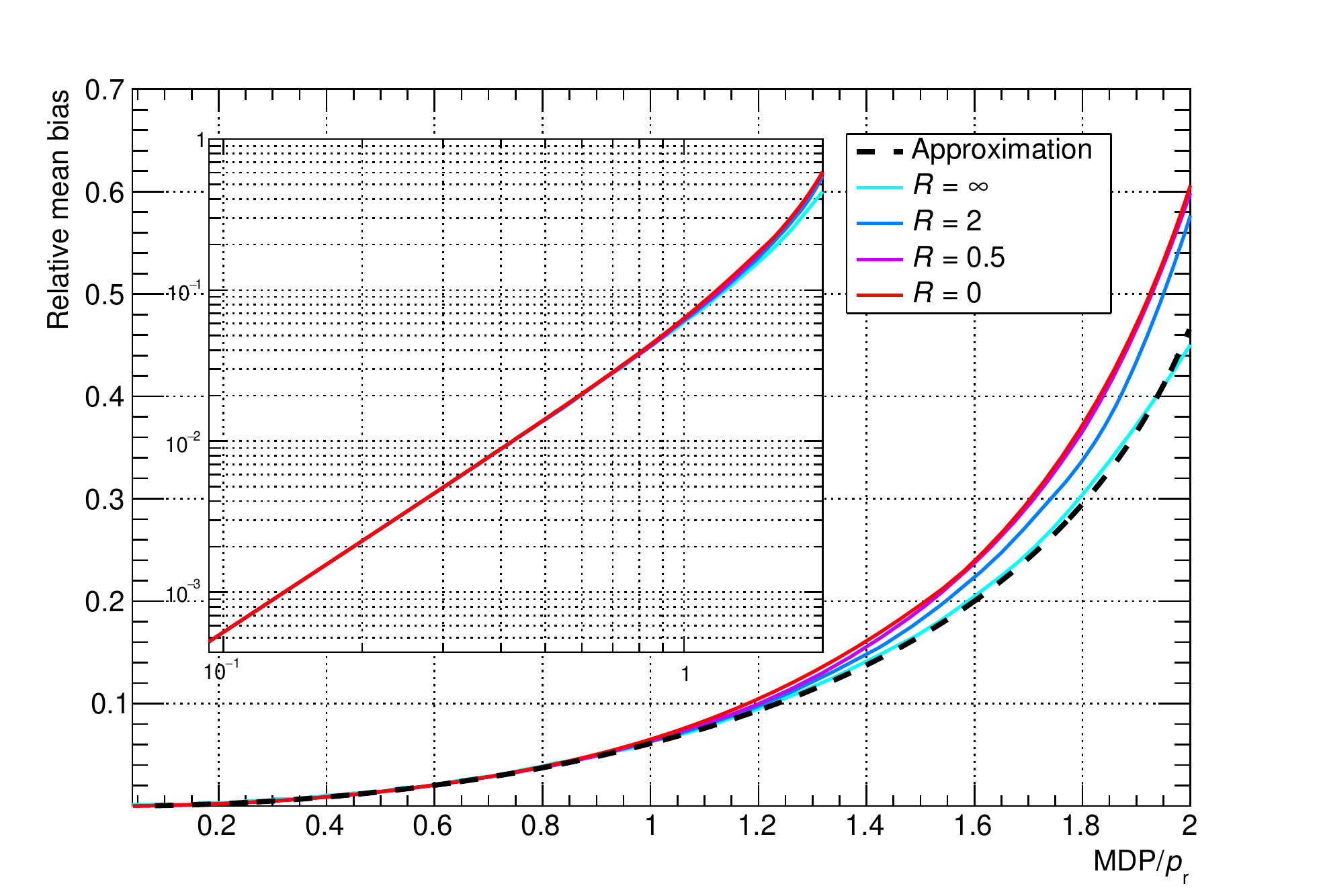}}
\caption{Relative mean bias $\beta$ as given by Eq. \ref{eq:beta_nummeric} (solid colored lines) and the approximation of Eq. \ref{eq:mean_bias_mdp_approx} (black dashed line). Here $p_0=\mu_0=1$ has been used to show the maximum difference between the different signal-to-background ratios $R$. A log-log plot is shown in the inset.}
\label{fig:mean_bias_SNR_r}
\end{figure}

Since $p_0$ is not known \textit{a priori}, Eq.~\ref{eq:mean_bias_mdp_approx_p0} needs to be expressed as a function of $p_{\mathrm{r}}$ by recursion. After some simplification the result is
\begin{equation}
\begin{split}
\beta\approx\frac{1-2x^2-\sqrt{1-4x^2}}{2x^2}
\end{split}
\label{eq:mean_bias_mdp_approx}
,\end{equation}
where $x\equiv\textrm{MDP}/4.29p_{\mathrm{r}}$. This shows that $\mathrm{MDP}/p_\mathrm{r}$ is a good choice of independent variable. Equation~\ref{eq:mean_bias_mdp_approx} is shown in Fig.~\ref{fig:mean_bias_SNR_r} where it is seen that $\beta>0$ and increases monotonously, i.e., on average the reconstructed polarization $p_{\mathrm{r}}$ will always be greater than the fixed polarization $p_0$. Exact numerical integration of Eq.~\ref{eq:beta_nummeric} is also provided for different signal-to-background ratios $R$ yielding similar results to the approximation in Eq.~\ref{eq:mean_bias_mdp_approx}. Here $R=0$ is the limit of large $N$, low $S$ and yet finite MDP. To avoid inaccurately computing $\beta$ for small $S$ (a problem under the CLT approximation) Eq.~\ref{eq:full_likelihood} is used for calculating $L(p_\mathrm{r}|p_0,\psi_0,\mu_0)$ and ultimately its mean when $S<200$.

To understand when the bias is significant with respect to the statistical uncertainty on $p_{\mathrm{r}}$, the bias fraction $(\mean{p_\mathrm{r}}-p_0)/\sigma_{p_\mathrm{r}}=\beta\times p_0/\sigma_{p_\mathrm{r}}$ is shown in Fig.~\ref{fig:sigma_r}. Here $\sigma_{p_\mathrm{r}}$ is derived (see Appendix~\ref{appendix_uncertainties}) using standard error propagation yielding
\begin{equation}
\sigma_{p_\mathrm{r}}=\frac{2}{\mu_\mathrm{r}}\sqrt{\frac{1}{S}\left(\frac{N}{2S}-\frac{\mu_0^2p_0^2}{4}\right)}
\label{eq:sigma_approx}
.\end{equation}
It becomes clear that this bias cannot be ignored in the low statistics regime. For a measured polarization below MDP, i.e., when $\textrm{MDP}/p_{\mathrm{r}}>1$, the bias is more than $18\%$ of the statistical uncertainty. Hence, one should not use Eq.~\ref{eq:define_p_r} for estimating $p_0$ in this regime.

\begin{figure}
\resizebox{\hsize}{!}{\includegraphics{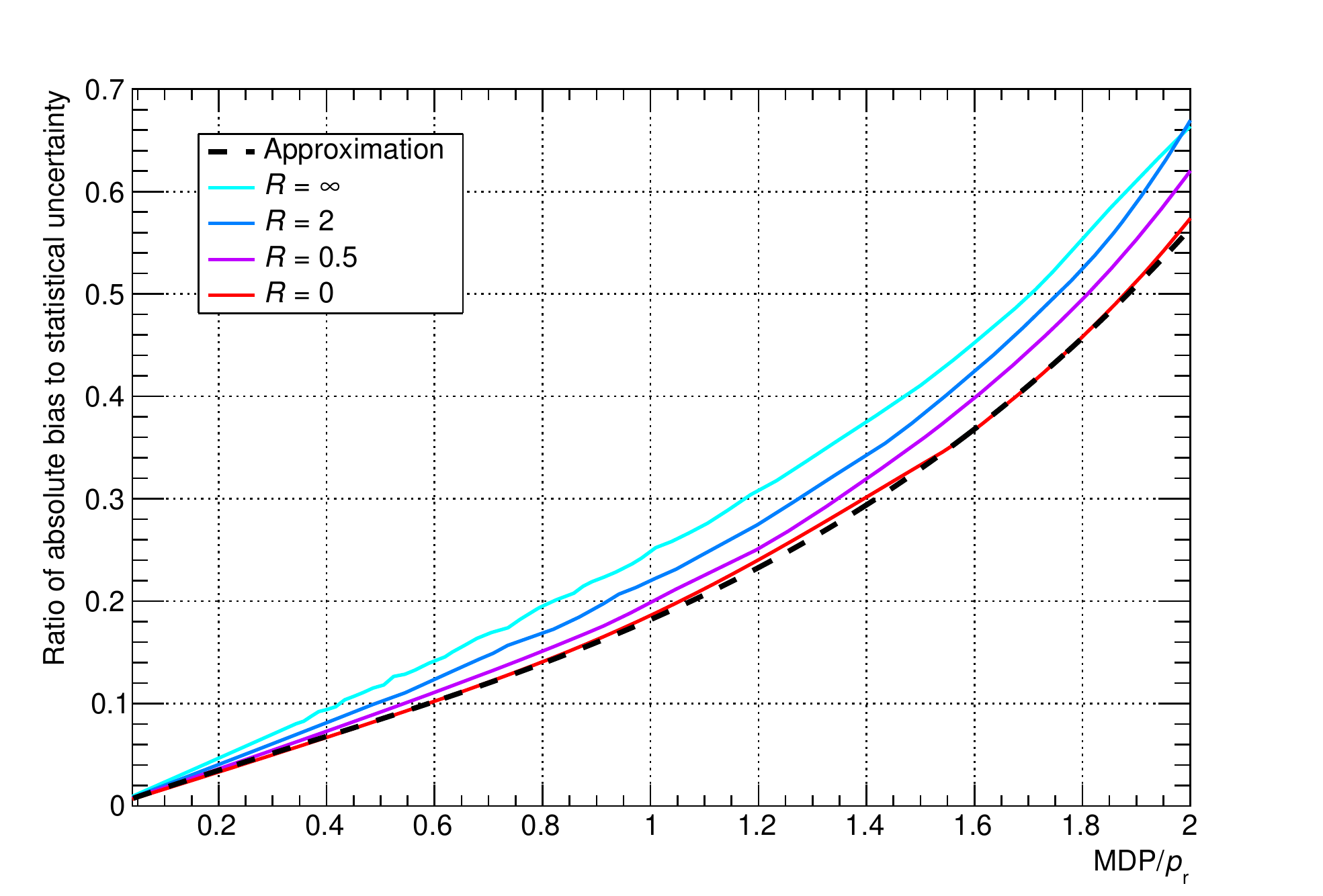}}
\caption{The ratio of the absolute bias to the statistical uncertainty $(\mean{p_\mathrm{r}}-p_0)/\sigma_{p_\mathrm{r}}=\beta\times p_0/\sigma_{p_\mathrm{r}}$. The $\beta$ in the approximation (black dashed line) is given by Eq.~\ref{eq:mean_bias_mdp_approx}. Here $p_0=\mu_0=1$ has been used to show the maximum difference between the different signal-to-background ratios $R$.}
\label{fig:sigma_r}
\end{figure}

A conclusion of this analysis is that significant errors are incurred when dividing the data into several data sets as is done for example by Dean et al. (\cite{Dean2008}) in order to estimate the statistical uncertainty. The smaller the data set, the bigger the bias, and thus, on average, the result of an integrated measurement will be lower than the mean of its constituent data sets. This is also relevant when fitting models to polarization fraction sub-divided with respect to energy or time. Binning the data will result in a higher reconstructed polarization fraction, thus biasing the fit and therefore physical conclusions should not be drawn from $p_\mathrm{r}$, as is done for example in Vadawale et al. (\cite{Vadawale2018}) where the evolution of polarization parameters throughout the Crab pulsar pulse phase is investigated.

\section{Maximum a posteriori estimate}

The previous section showed the bias arising when using Eq.~\ref{eq:define_p_r} for estimating $p_0$ independently of the polarization angle. It is not obvious if such a bias exists when using $(p_\mathrm{r},\psi_r)$ for the joint point estimate of $(p_0,\psi_0)$. However, in this case it does not make sense to find a $\mean{p_\mathrm{r},\psi_r}$ for a fixed $(p_0,\psi_0)$ so here the word "bias" is interpreted as the difference between the Maximum A Posteriori (MAP) estimate, $p_\mathrm{MAP}$, and $p_\mathrm{r}$. Such an analysis requires a Bayesian approach.

\subsection{Central Limit Theorem approximation of the posterior}

The posterior $P(p_0,\Delta\psi_0|p_{\mathrm{r}})$ is derived by applying the Bayes theorem to the likelihood given by Eq.~\ref{eq:likelihood_polar}
\begin{equation}
P(p_0,\Delta\psi_0|p_{\mathrm{r}})=\mathcal{N}\times P(p_0,\Delta\psi_0) \times L(p_{\mathrm{r}},0|p_0,\Delta\psi_0)
\label{eq:posterior}
,\end{equation}
where $\mathcal{N}$ is the normalization factor, $P(p_0,\psi_0)$ is the prior and $\Delta\psi_0=\psi_0-\psi_r$. Since in a Bayesian approach the parameters $(p_0,\psi_0)$ vary and the data $(p_\mathrm{r},\psi_\mathrm{r})$ are fixed, $L(p_{\mathrm{r}},0|p_0,\Delta\psi_0)$ is a function of $\psi_0$. It follows that $\sigma_Q\neq\sigma_U$ and $\rho\neq0$. This differs from the typical assumption of $\sigma_Q=\sigma_U$ and no correlation between $Q$ and $U$ as done in other works (Simmons \& Stewart~\cite{Simmons1985}; Vaillancourt~\cite{Vaillancourt2006}; Quinn~\cite{Quinn2012}; Maier et al.~\cite{Maier2014}).

The Jeffreys prior has the desirable property of being invariant under re-parametrization (Jeffreys~\cite{Jeffreys1939}). In this case it is a uniform prior in the Cartesian coordinates $(Q_0,U_0)$. It will result in a preference of high polarization fractions after transforming to polar coordinates, as discussed in Quinn (\cite{Quinn2012}). This is unphysical because it is more difficult to make a highly polarized photon beam in nature since any disorder in the emission region will lower the polarization fraction. A more realistic approach is to instead take a uniform prior $0\le p_0 \le 1$ in polar coordinates. The prior for $\psi_0$ is also uniform due to symmetry.

\begin{figure}
\resizebox{\hsize}{!}{\includegraphics{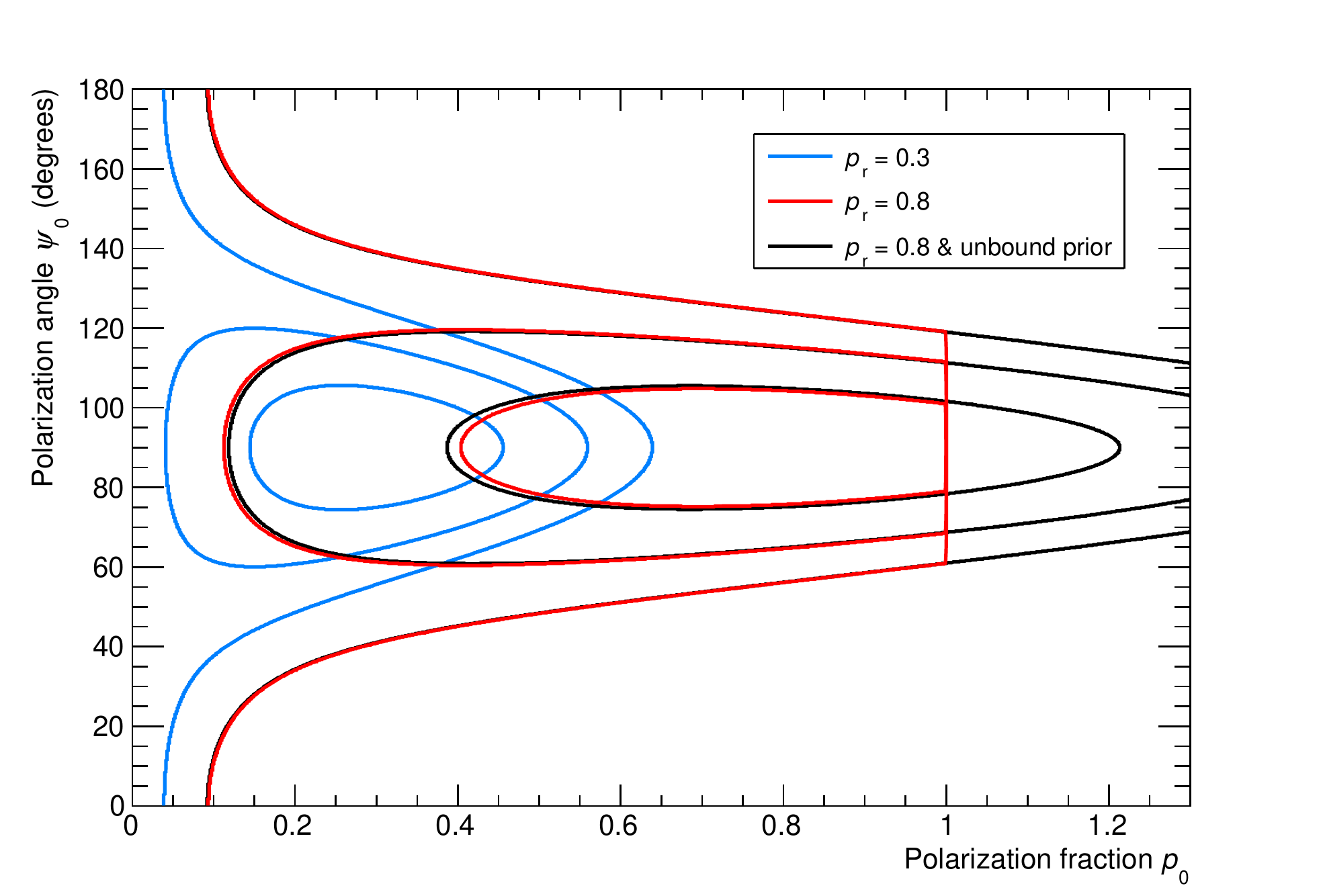}}
\caption{The highest posterior density credibility regions for the posterior given by Eq.~\ref{eq:posterior} for $\mathrm{MDP}/p_\mathrm{r}=1$, $R=0$ and $\mu_0=1$. The contours correspond to $1\sigma$, $2\sigma$ and $3\sigma$ probability content. The posterior is truncated by the prior at $p_0=1$. The unbound prior scenario (black line) corresponds to the unnormalized prior $p_0\geq0$. }
\label{fig:posterior_2d}
\end{figure}

Figure~\ref{fig:posterior_2d} shows the posterior as given by Eq.~\ref{eq:posterior} for a low-statistics measurement $\mathrm{MDP}/p_\mathrm{r}=1$. The asymmetry is apparent as the posterior broadens for low values of $p_0$. The effect of the prior is illustrated by including the unbound prior scenario where $p_0\geq0$.

\begin{figure}
\resizebox{\hsize}{!}{\includegraphics{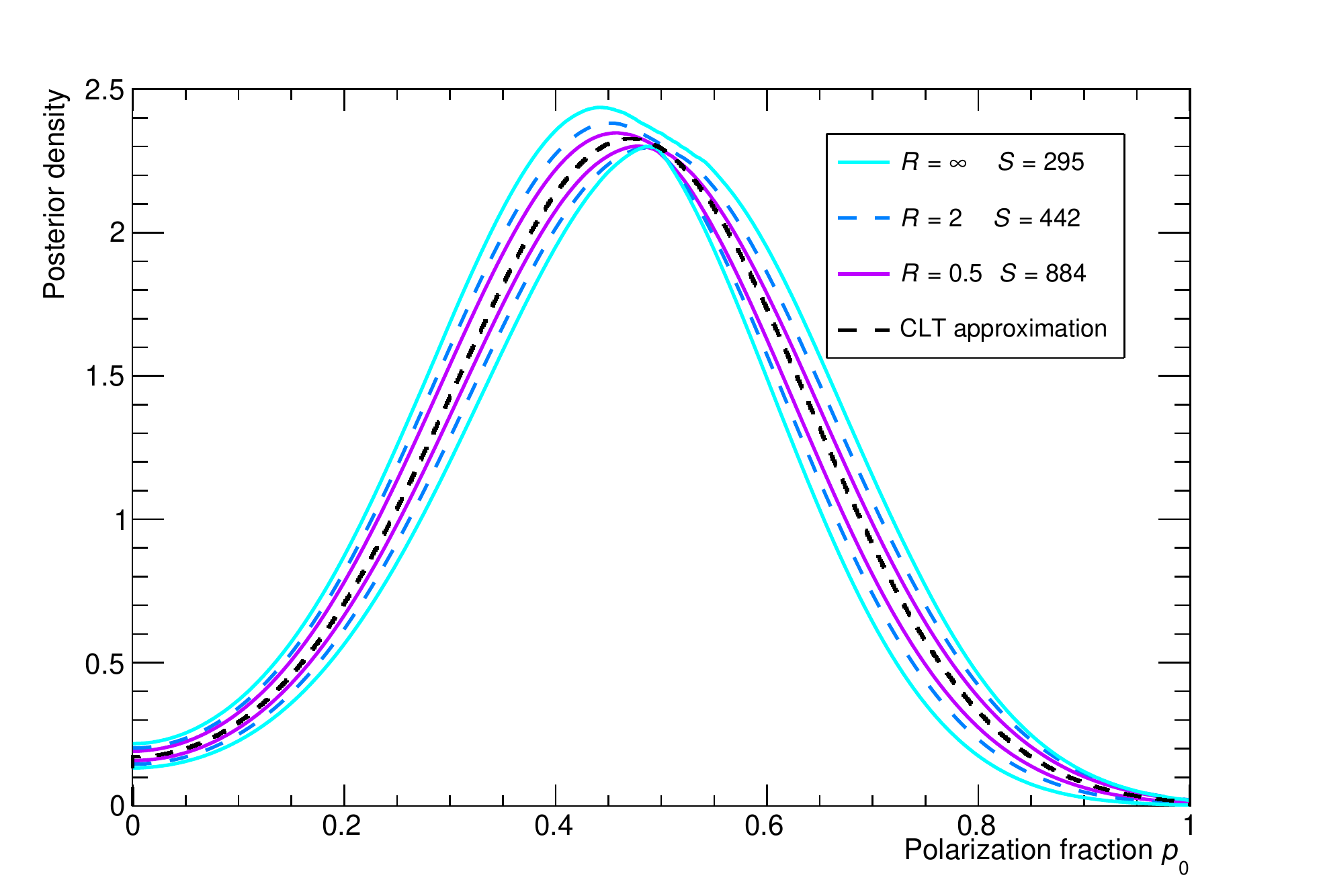}}
\caption{The intervals containing 90\% of the posteriors for $(p_\mathrm{r}=0.5,\psi_\mathrm{r}=22.5^\circ)$ and the CLT approximation. As the number of signal photons increases, the posteriors converge towards the approximation. Here $\mu_0=\mu_\mathrm{r}=0.5$ and $\mathrm{MDP}/p_\mathrm{r}=1$.}
\label{fig:posterior_density_0_5_prior}
\end{figure}

Since the posterior is derived using a likelihood which utilizes the CLT approximation, the posterior may, for certain combinations of parameters, be inaccurate. To check the magnitude of the effect, pairs of $(p_0,\psi_0)$ are sampled from the prior and then used to generate a data set. Data sets falling within a narrow window of a chosen $p_\mathrm{r}$ and $\psi_\mathrm{r}$ ($p_\mathrm{r}\pm0.002$ and $\psi_\mathrm{r}\pm0.36^\circ$) are selected and the posterior for each data set is found using Eq.~\ref{eq:full_likelihood} as the likelihood. Finally, the intervals containing 90\% of the posteriors marginalized over the polarization angle (for clarity) are shown in Fig.~\ref{fig:posterior_density_0_5_prior} along with the CLT approximation. The situation is similar for the posterior of the polarization angle marginalized over the polarization fraction.

The most important feature of Fig.~\ref{fig:posterior_density_0_5_prior} is that the posterior is not the same for fixed $(p_\mathrm{r},\psi_\mathrm{r})$ when the number of signal photons is low. Hence $(p_\mathrm{r},\psi_\mathrm{r})$ are not always sufficient statistics and the full data set $\{\psi_i\}$ is required for deriving the posterior. However, the posterior converges quickly towards the CLT approximation when $R<0.5$ since any meaningful measurement with such a low $R$ requires a large number of photons. An effect not shown in the figure is that the difference between the full posterior and the CLT approximation increases as $p_\mathrm{r}$ and $\mu_\mathrm{r}$ increase requiring more photons for good convergence. However, few X-ray polarimeters have $\mu_0$ higher than 0.5 (Krawczynski et al.~\cite{Krawczynski2011}; Kaaret ~\cite{Kaaret2014}) and synchrotron emission, a proposed mechanism for many astrophysical sources, is not expected to produce a polarization fraction above $\sim0.6$ (Lyutikov et al. \cite{Lyutikov2003}).

In conclusion, the full likelihood in Eq.~\ref{eq:full_likelihood} rather than the CLT approximation in Eq.~\ref{eq:likelihood_polar} is required for sources where the signal-to-background ratio is high but the total number of photons is low (e.g., short-duration GRBs). A guideline is to use the CLT approximation only for data sets where $S>10^3$. As mentioned before, polarimeters foreseen to be active in the near future (Astrosat/CZTI, POLAR, IXPE, X-Calibur) have $\mu_0<0.5$ so this guideline translates to the CLT approximation being generally valid for \mbox{$\mathrm{MDP}<0.27$}.

\subsection{Relative MAP bias}

The MAP estimate is at $p_{_{\mathrm{MAP}}}> p_{\mathrm{r}}$ and $\psi_0=\psi_{\mathrm{r}}$. The inequality occurs because $\sigma_Q$ and $\sigma_U$ depend on $p_0$ but only measurements with high $R$ and $p_0^2\mu_0^2\gg0$ are significantly affected. In all other cases $\sigma_Q\approx\sigma_U\approx\sqrt{2N}/\mu_0 S$ and $p_{_\mathrm{MAP}}\approx p_{\mathrm{r}}$. 

The relative MAP bias is defined as
\begin{equation}
\beta_{_\mathrm{MAP}}\equiv\frac{p_{\mathrm{r}}-p_{_\mathrm{MAP}}}{p_{_\mathrm{MAP}}}
\label{eq:map_bias}
.\end{equation}
Table~\ref{tab:map_bias} shows $\beta_{_\mathrm{MAP}}$ for the extreme case $\mathrm{MDP}/p_{\mathrm{r}}=2$ (a monotonously decreasing function) for different polarization parameters. Only scenarios with a sufficient number ($S>884$ based on the results shown in Fig.~\ref{fig:posterior_density_0_5_prior}) of signal photons are considered. The results show that $\beta_{_\mathrm{MAP}}$ increases with $p_\mathrm{r}$ and $\mu_0$ but its value is negligible for all cases where the CLT approximation is valid. Therefore, when $(p_\mathrm{r},\psi_\mathrm{r})$ are sufficient statistics of the data, they are a good estimate for the MAP.

\begin{table}
\caption{The relative MAP bias $\beta_\mathrm{MAP}$ for $\mathrm{MDP}/p_\mathrm{r}=2$, different signal-to-background ratios $R$, sufficient statistic $p_\mathrm{r}$ and modulation factor $\mu_0$.}
\label{tab:map_bias}
\centering
\begin{tabular}{|c|c|c|c|}
\hline
$R$ & $p_\mathrm{r}$ & $\mu_0$ & $\beta_\mathrm{MAP}$ \\
\hline
\hline
0 & $0.5$ & $0.5$ & $-8\times10^{-6}$ \\
\hline
0 & $0.8$ & $1.0$ & $-3\times10^{-4}$ \\
\hline
0.5 & $0.5$ & $0.25$ & $-7\times10^{-5}$ \\
\hline
\end{tabular}
\end{table}

The broadening of the joint posterior at low values of $p_0$, as shown in Fig.~\ref{fig:posterior_2d}, results in $p_{_\mathrm{MAP}}$ being a poor estimate for $p_0$ when marginalizing over $\psi_0$. More generally, this occurs for any two-dimensional distribution when marginalizing over the nuisance parameter if the estimated parameter has an asymmetric distribution.

\section{Uncertainties on polarization parameters}

There are two reasons why it is not always appropriate to use naive Stokes estimates for uncertainties of polarization parameters: non-Gaussianity and the prior. Eq.~\ref{eq:sigma_approx} gives the correct uncertainty on the polarization fraction when statistics are high and the reconstructed fraction $p_{\mathrm{r}}$ is well below 1. If these conditions are not fulfilled, the marginalized posterior becomes highly asymmetric either due to low statistics or because a part of the likelihood is truncated by the prior at $p_0=1$ and is thus poorly approximated by a Gaussian.

\begin{figure}
\resizebox{\hsize}{!}{\includegraphics{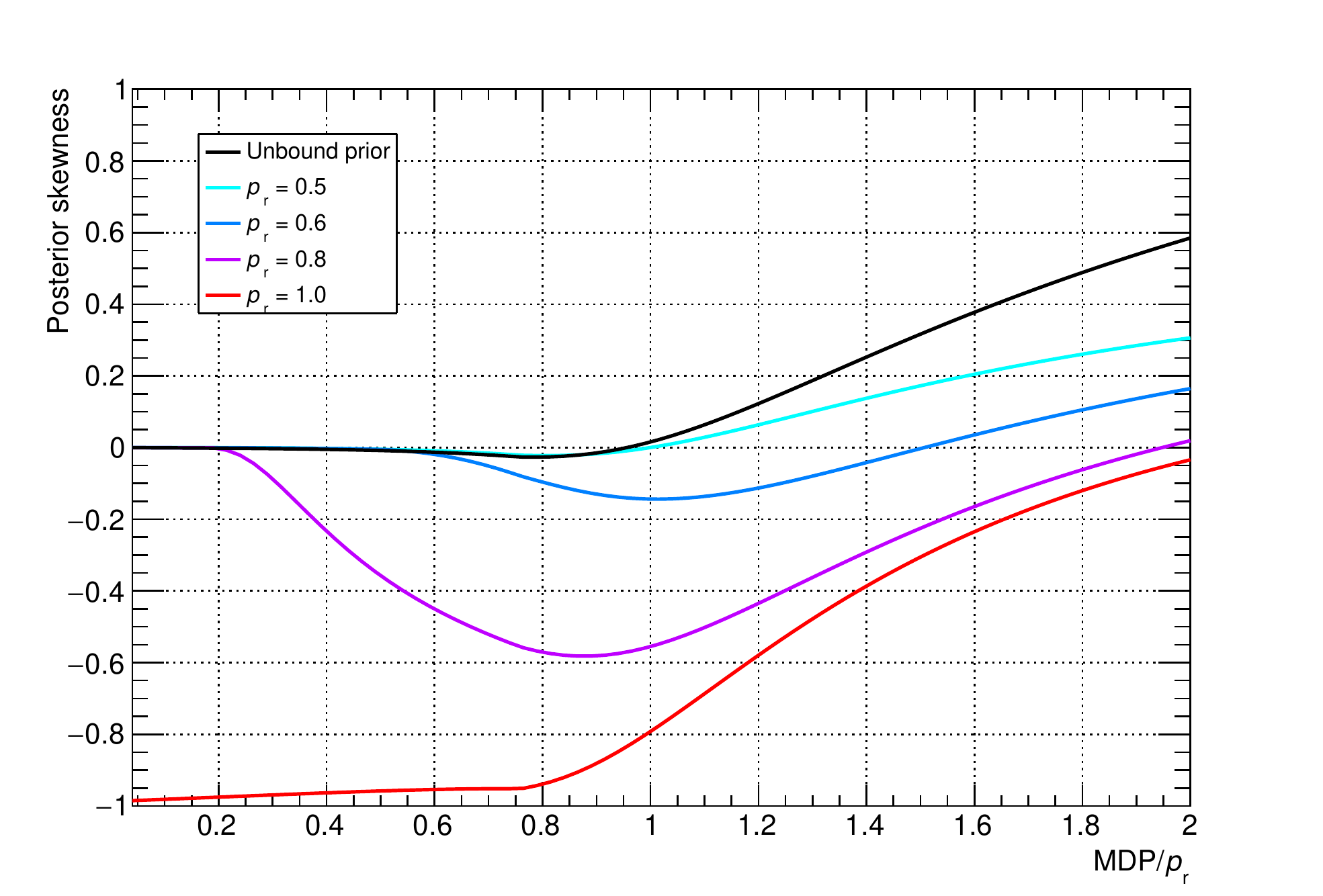}}
\caption{Skewness of the marginalized posterior for the polarization fraction. The unbound prior scenario (black line) uses $p_0\geq0$ and $p_{\mathrm{r}}=1$ (here the result is independent of $p_r$). A positive skewness implies an asymmetric distribution with a longer tail on the right, while a negative skewness results in a tail on the left. In all cases $R=0$ is assumed.}
\label{fig:skewness}
\end{figure}

The asymmetry of the marginalized posterior for the polarization fraction is shown in terms of the skewness in Fig.~\ref{fig:skewness}. The unbound prior scenario employs the unnormalized uniform prior $p_0\geq0$ and $p_{\mathrm{r}}=1$ (here all values of $p_{\mathrm{r}}$ produce the same result). It shows that when statistics are low, the marginalized posterior gets a longer tail on the right (positive skewness). This is because the peak moves left, which is equivalent to measuring a higher $p_\mathrm{r}$ for a fixed $p_0$, that is, the same interpretation as for the relative mean bias shown in Fig.~\ref{fig:mean_bias_SNR_r}.

Although the marginalized posterior for the polarization angle is always symmetric, it deviates from Gaussianity when statistics are low, for example, approaching the uniform distribution since all angles are equally likely. It is also affected by the prior when $p_{\mathrm{r}}$ is sufficiently high, resulting in a distribution with longer tails because, as shown in Fig.~\ref{fig:posterior_2d}, the prior only truncates the part of the likelihood contributing to the central part of the marginalized distribution. Therefore, the uncertainty given by
\begin{equation}
\sigma_{\psi_{\mathrm{r}}}=\frac{1}{\sqrt{2}\mu_\mathrm{r} p_{\mathrm{r}}}\frac{\sqrt{N}}{S}
\label{eq:sigma_angle}
\end{equation}
(derived in Appendix~\ref{appendix_uncertainties}) is not always valid.

In a Bayesian approach, the posterior is used instead of a parameter estimate and its uncertainty since the posterior contains the full information about the parameter. The posterior cannot easily be described in text so a simplified description is necessary such as its peak and the region of Highest Posterior Density (HPD) containing the probability content corresponding to one Gaussian standard deviation.

\begin{figure}
\resizebox{\hsize}{!}{\includegraphics{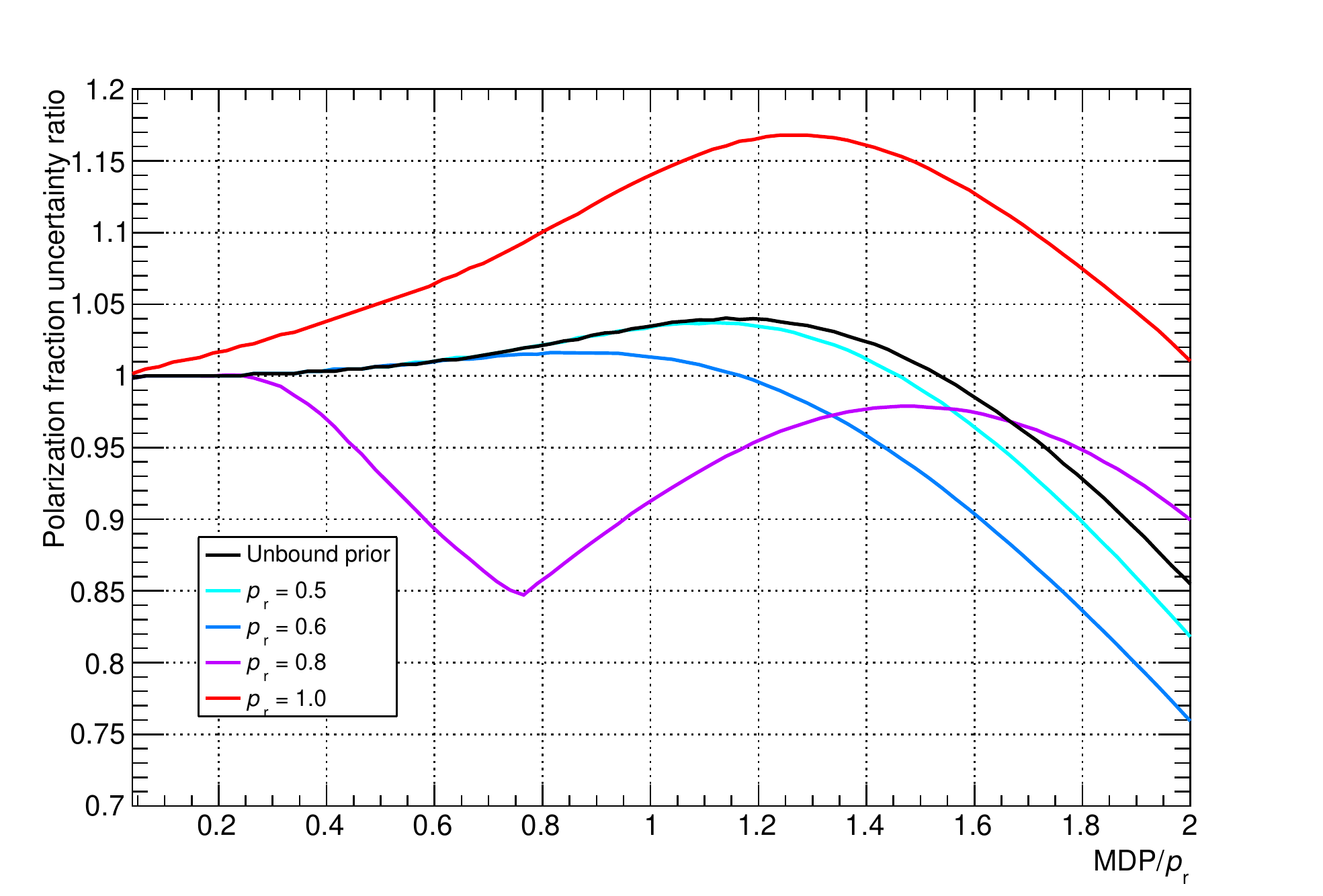}}
\caption{The ratio $\sigma_{p_{_\mathrm{HPD}}}/\sigma_{p_{\mathrm{r}}}$ of the uncertainty derived from the HPD region of the marginalized posterior and the Gaussian uncertainty given by Eq.~\ref{eq:sigma_approx} for the polarization fraction. Since the posterior is asymmetric, an effective  $\sigma_{p_{_\mathrm{HPD}}}$ is defined as half of the region containing $1\sigma$ Gaussian probability. The unbound prior scenario corresponds to the unnormalized prior $0\leq p_0$ and $p_{\mathrm{r}}=1$ (here the result is independent of $p_r$). In all cases $R=0$ is assumed.}
\label{fig:sigma_ratio}
\end{figure}

\begin{figure}
\resizebox{\hsize}{!}{\includegraphics{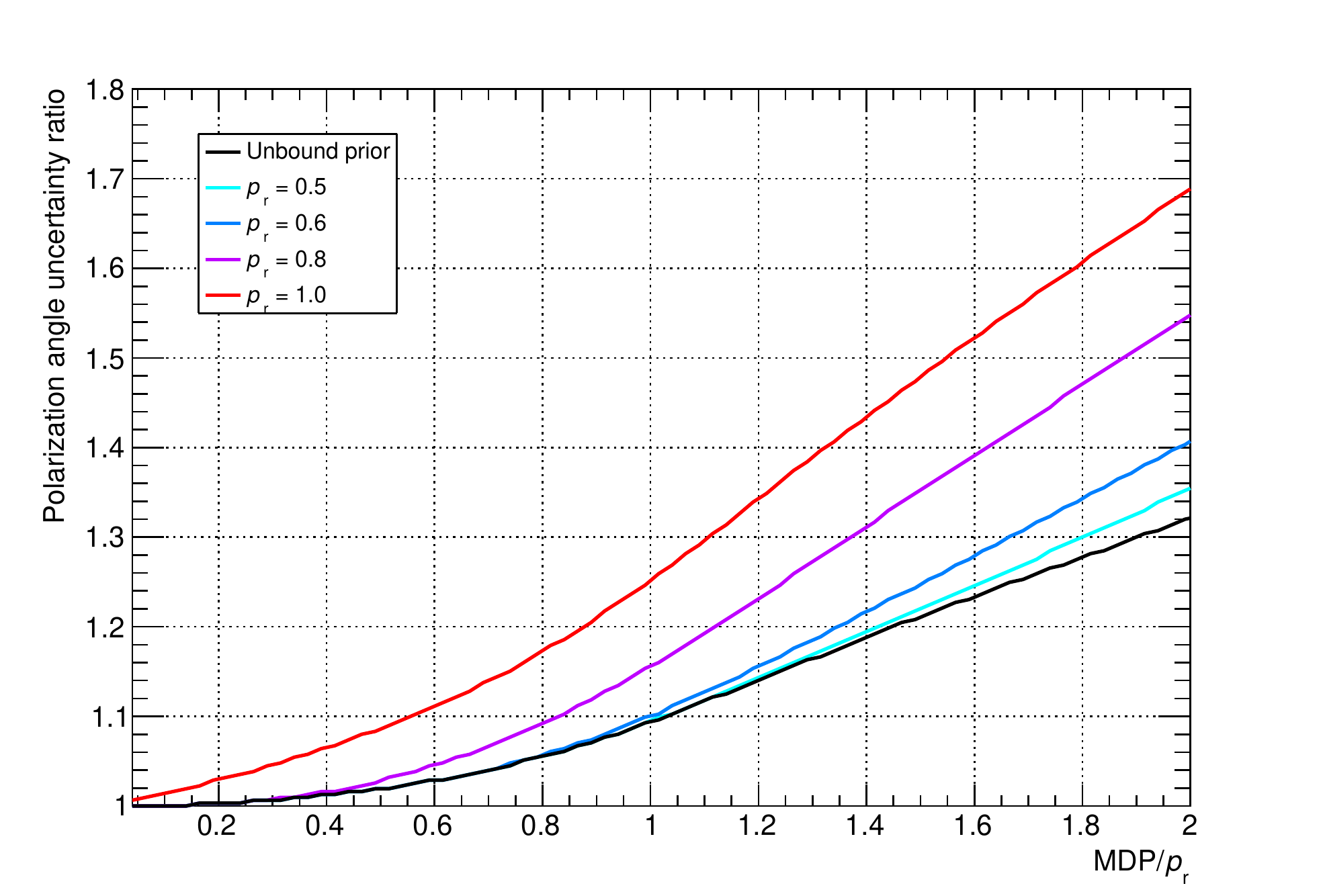}}
\caption{The ratio $\sigma_{\psi_{_\mathrm{HPD}}}/\sigma_{\psi_{\mathrm{r}}}$ of the uncertainty derived from the HPD region of the marginalized posterior and the Gaussian uncertainty given by Eq.~\ref{eq:sigma_angle} for the polarization angle. The unbound prior scenario corresponds to the unnormalized prior $0\leq p_0$ and $p_{\mathrm{r}}=1$ (here the result is independent of $p_r$). In all cases $R=0$ is assumed.}
\label{fig:sigma_angle_ratio}
\end{figure}

It is possible to quantify when the HPD region must be derived from the posterior and when Eq.~\ref{eq:sigma_approx} and Eq.~\ref{eq:sigma_angle} are good approximations by considering the ratios $\sigma_{p_{_\mathrm{HPD}}}/\sigma_{p_{\mathrm{r}}}$ and $\sigma_{\psi_{_\mathrm{HPD}}}/\sigma_{\psi_{\mathrm{r}}}$ as functions of $\mathrm{MDP}/p_{\mathrm{r}}$, which are shown in Figs.~\ref{fig:sigma_ratio} and \ref{fig:sigma_angle_ratio},  respectively. In both figures, the unbound prior scenario uses the unnormalized uniform prior $p_0\geq0$ and $p_{\mathrm{r}}=1$. Only scenarios with $R=0$ are considered, meaning that the CLT approximation is valid in the entire range of $\mathrm{MDP}/p_{\mathrm{r}}$ for any $p_\mathrm{r}$. Increasing $R$ does not change the results shown in this section but may invalidate the CLT approximation. Such cases cannot be represented because $p_\mathrm{r}$ is not a sufficient statistic. However, since the posterior shape does not change drastically (as shown in Fig.~\ref{fig:posterior_density_0_5_prior}), the following discussion is a good qualitative description of its behavior for any $R$.

For ease of comparison, $\sigma_{p_{_\mathrm{HPD}}}$ is defined as half of the region containing $1\sigma$ Gaussian probability. Additionally, the Gaussian interval is limited to account for the prior, for example, $p_0=0.8\pm0.3$ is the interval $[0.5,1.0]$ which gives an uncertainty of $\sigma_{p_{\mathrm{r}}}=0.25$,  and therefore it does not contain 68.3\% ($1\sigma$) probability content. 

The unbound prior scenario in Fig.~\ref{fig:sigma_ratio} shows that for $\mathrm{MDP}/p_{\mathrm{r}}\sim1$ the marginalized posterior has longer tails than a Gaussian. Additionally, the prior truncates the posterior for high $p_\mathrm{r}$ or for low statistics. This can result in either a higher or a lower $\sigma_{p_{_\mathrm{HPD}}}/\sigma_{p_{\mathrm{r}}}$ depending on where exactly the distribution is truncated. Unless $p_{\mathrm{r}}$ is high, $\sigma_{p_{\mathrm{r}}}$ is a good approximation for the magnitude of the uncertainty at $\mathrm{MDP}/p_{\mathrm{r}}\sim1$ but it is important to remember that the posterior is asymmetric for such low statistics.

For the polarization angle, Fig.~\ref{fig:sigma_angle_ratio} shows that $\sigma_{\psi_{_\mathrm{HPD}}}>\sigma_{\psi_{\mathrm{r}}}$. The situation is simpler than for $p_0$ because the angle has a symmetric posterior. The unbound prior shows the deviation from a Gaussian manifesting in longer tails. Figure~\ref{fig:posterior_2d} shows that when adding the prior $0\leq p_0 \leq 1$, the part of the likelihood which extends past $p_0=1$ is truncated. This part contributes to the density at the center of the marginalized posterior. Removing it further extends the tails. It follows that Eq.~\ref{eq:sigma_angle} underestimates the uncertainty on the polarization angle by relative $10\%-20\%$ at $\mathrm{MDP}/p_{\mathrm{r}}\sim1$.

This analysis shows that in the limit of low statistics, $\mathrm{MDP}/p_\mathrm{r}\sim1$, the uncertainty on the polarization angle is not well-described by Eq.~\ref{eq:sigma_angle} and a Bayesian treatment is necessary. For high $p_r$, such a treatment is required even for high-statistics measurements, $\mathrm{MDP}/p_\mathrm{r}\sim0.5$, because of the asymmetry in the uncertainty on the polarization fraction.

To illustrate the effects described above, two examples are presented in Table~\ref{tab:example}. The first is a recent measurement of the Crab nebula by PoGO+ (Chauvin et al.~\cite{Chauvin2017PoGO+}) and the second is a hypothetical measurement of a high reconstructed polarization fraction highlighting the importance of the prior $0\leq p_0\leq 1$. The largest differences are in the polarization fraction and the uncertainty on the polarization angle.

\begin{table}
\caption{Examples of the difference between a Bayesian approach and a Gaussian approximation. The first row is a recent measurement of the Crab nebula by PoGO+ (Chauvin et al.~\cite{Chauvin2017PoGO+}). The second row is a hypothetical measurement of a high reconstructed polarization fraction.}
\label{tab:example}
\centering
\begin{tabular}{|c|c|c|c|c|}
\hline
$\mathrm{MDP}/p_\mathrm{r}$ & $p_\mathrm{r}$ (\%) & $p_0$ (\%) & $\psi_\mathrm{r}$ & $\psi_0$ \\
\hline
\hline
$1.3$ & $19.5\pm8.3$ & $17.4^{+8.6}_{-9.3}$ & $137\pm12^\circ$ & $137\pm15^\circ$ \\
\hline
$1$ & $80\pm26$ & $76^{+21}_{-22}$ & $90\pm9^\circ$ & $90\pm11^\circ$ \\
\hline
\end{tabular}
\end{table}

\section{Uncertainties on the modulation factor}

The uncertainty $\sigma_\mu$ on the modulation factor $\mu_0$ can be minimized for most polarimeters designed for measuring point sources by increasing the statistics during calibration tests. However, for polarimeters measuring GRBs it is often impossible to make $\sigma_\mu$ arbitrarily small. The problem is that $\mu$ varies depending on the location of the GRB, typically having the highest values for on-axis GRBs but significantly lower values for GRBs located at a large angular separation from the detector axis. Determining the location of the GRB by using a polarimeter involves large uncertainties since it is often not optimized for the task. The uncertainty on the location propagates to a non-negligible $\sigma_\mu$. Additionally, the primary spectrum of GRBs may not be reconstructed with sufficient precision by the polarimeter, introducing further uncertainties in the simulation required for deducing the $\mu_0$ for a particular GRB. If the GRB is simultaneously observed by a dedicated GRB monitor, the uncertainty on its location and spectrum will be smaller, but no GRB monitor has complete sky coverage or 100\% duty-cycle, so it is inevitable that some GRBs will only be seen by the polarimeter.

An example of a GRB polarimeter is POLAR (Produit et al. \cite{Produit2005}). For a typical bright GRB,  POLAR is expected to have a $\sigma_\mu/\mu$ of between 5 and 15\% assuming that the burst occurs on-axis (Suarez-Garcia et al. \cite{Suarez-Garcia2010}). The simplest way to account for this additional uncertainty is to propagate $\sigma_\mu$ through

\begin{equation}
p_{\mathrm{r}}=\frac{M}{\mu_{\mathrm{r}}}
,\end{equation}
where $M$ is the measured modulation. The total symmetric uncertainty is then given by
\begin{equation}
\sigma_\mathrm{tot}=p_\mathrm{r}\sqrt{\bigg(\frac{\sigma_\mu}{\mu_\mathrm{r}}\bigg)^2+\bigg(\frac{\mu_\mathrm{r}\sigma_{p_\mathrm{r}}}{M}\bigg)^2}
\label{eq:sigma_tot}
.\end{equation}

\begin{figure}
\resizebox{\hsize}{!}{\includegraphics{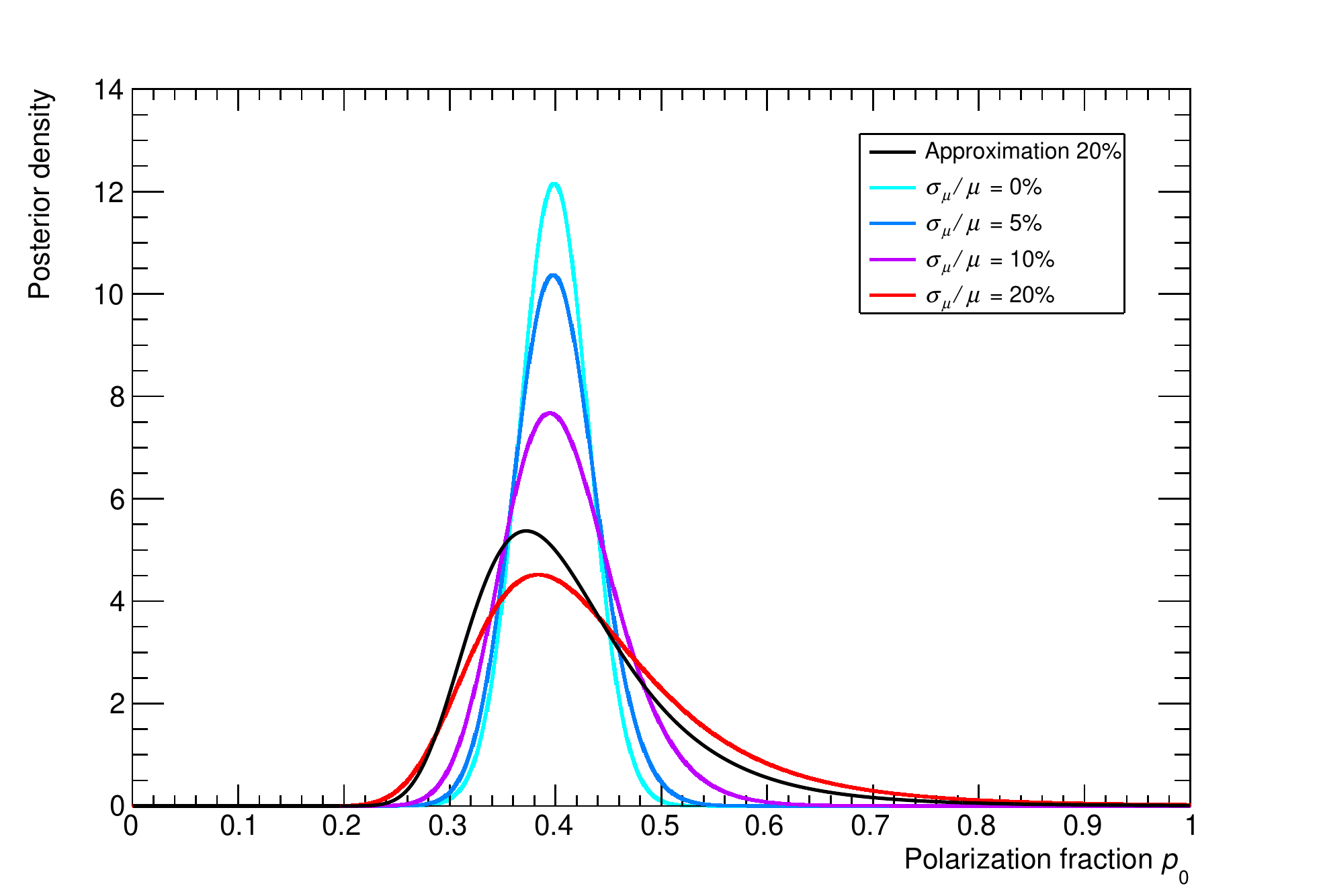}}
\caption{Polarization fraction posterior density for $\mathrm{MDP}/p_{\mathrm{r}}=0.25$, $p_{\mathrm{r}}=0.4$ and $\mu_{\mathrm{r}}=0.4$ for different $\sigma_\mu/\mu$. As $\sigma_\mu/\mu$ increases, the distribution becomes wider and more asymmetric. The approximation is the reciprocal Gaussian distribution in Eq.~\ref{eq:rec_gaus}.}
\label{fig:posterior_m100}
\end{figure}

To check for which parameters the symmetry is a good approximation, an additional prior can be added to the posterior Eq.~\ref{eq:posterior}. For simplicity, this prior is assumed to be Gaussian but it can vary depending on localization sensitivity. The nuisance parameter $\mu_0$ can then be marginalized over, yielding
\begin{equation}
\begin{split}
P(p_0,\Delta\psi_0|p_{\mathrm{r}},\mu_{\mathrm{r}})= \mathcal{N}\int^1_0 P(p_0,\Delta\psi_0)
\times \\
\exp{\bigg[-\frac{(\mu_{\mathrm{r}}-\mu_0)^2}{2\sigma_\mu^2}\bigg]}\times L(p_{\mathrm{r}},0,\mu_\mathrm{r}|p_0,\Delta\psi_0)
 \mathrm{d}\mu_0
\label{eq:posterior_m100}
\end{split}
,\end{equation}
where the integral is taken between 0 and 1 because $\mu_0>1$ is unphysical. As an example, the posterior (marginalized over the polarization angle) for a signal-only measurement at MDP=10\% (CLT approximation is valid because $S$ is large), $p_\mathrm{r}=0.4$ and $\mu=0.4$ is shown in Fig.~\ref{fig:posterior_m100} for different $\sigma_\mu/\mu$. Due to the high statistics, $\mathrm{MDP}/p_{\mathrm{r}}=0.25$, the distribution is symmetric for low $\sigma_\mu/\mu$, however, the tail on the right grows rapidly as $\sigma_\mu/\mu$ increases. The behavior is governed by the reciprocal Gaussian distribution which the posterior approaches in the limit of high photon statistics

\begin{equation}
g(p_0|p_{\mathrm{r}},\mu_{\mathrm{r}})=\frac{1}{\sqrt{2\pi}\sigma_\mu}\frac{p_{\mathrm{r}}\mu_{\mathrm{r}}}{p_0^2}\exp{\bigg[-\frac{(p_{\mathrm{r}}\mu_{\mathrm{r}}/p_0-\mu_{\mathrm{r}})^2}{2\sigma_\mu^2}\bigg]}
\label{eq:rec_gaus}
,\end{equation}
shown as an approximation in Fig.~\ref{fig:posterior_m100}. If not for the prior $0\le p_0 \le1$, the moments of this distribution would be undefined.

Figure~\ref{fig:posterior_m100} demonstrates that for $\sigma_\mu/\mu>10\%$ the shape of the posterior changes significantly and Eq.~\ref{eq:posterior_m100} is required. In the extreme case of $\sigma_\mu/\mu=20\%$, the half-width of the HPD region is 0.092, whereas Eq.~\ref{eq:sigma_tot} yields 0.086, implying not only that the uncertainty is asymmetric but also that it is significantly larger.

\begin{figure}
\resizebox{\hsize}{!}{\includegraphics{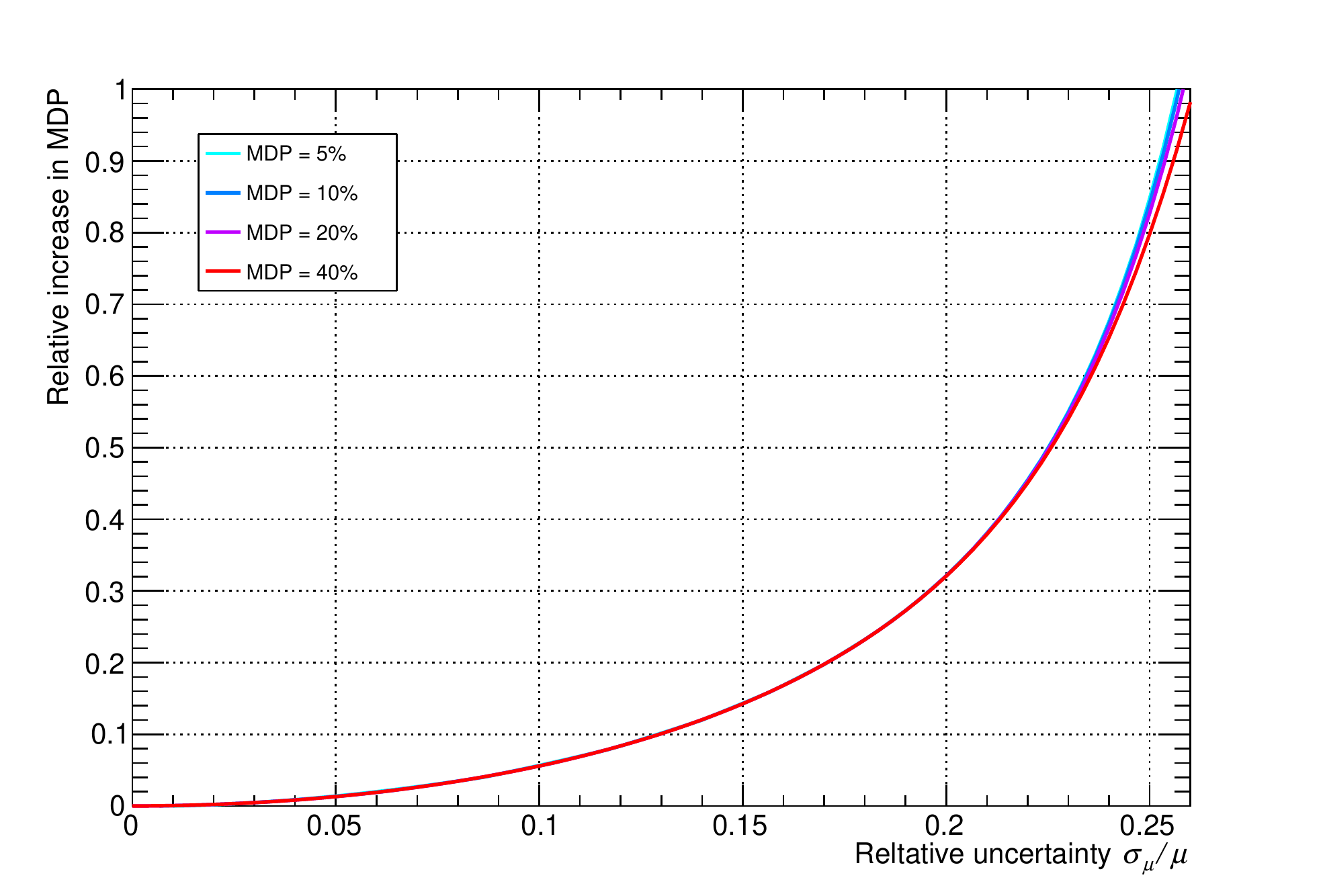}}
\caption{The relative increase in MDP, defined as the ratio $\mathrm{MDP}_\sigma/\mathrm{MDP}$ where $\mathrm{MDP}_\sigma$ is the solution to $\int_0^\mathrm{MDP_\sigma}f(p_{\mathrm{r}}|p_0=0,\mu_0)\mathrm{d}{p_\mathrm{r}}=0.99$ as a function of the relative uncertainty on $\mu$. The results are independent of the initial MDP.}
\label{fig:mdp_m100}
\end{figure}

These results can be related to the instrument perfomance by studying the effect $\sigma_\mu/\mu$ has on the MDP. A frequentist approach must be followed since MDP is a frequentist concept. MDP is derived by finding the 99\% upper limit for the Rayleigh distribution -- a special case of the Rice distribution where $p_0=0$. However, since there is an uncertainty on $\mu,$ the Rayleigh distribution must be multiplied by the likelihood for $\mu$ (assumed to be a Gaussian for simplicity) and integrated to yield 
\begin{equation}
\begin{split}
f(p_{\mathrm{r}}|p_0=0,\mu_0)=\int_0^\infty \frac{p_{\mathrm{r}}\mu_{\mathrm{r}}}{\sqrt{2\pi}\pi\sigma^2\sigma_\mu} \\
\times\exp{\bigg[-\frac{\mu_{\mathrm{r}}^2 p_{\mathrm{r}}^2}{2\sigma^2}-\frac{(\mu_{\mathrm{r}}-\mu_0)^2}{2\sigma_\mu^2}\bigg]}\mathrm{d}\mu_{\mathrm{r}}
\end{split}
.\end{equation}
Finally, $f(p_{\mathrm{r}}|p_0=0,\mu_0)$ can be integrated to find the 99\% upper limit so that $\int_0^\mathrm{MDP_\sigma}f(p_{\mathrm{r}}|p_0=0,\mu_0)\mathrm{d}p_\mathrm{r}=0.99$. Figure~\ref{fig:mdp_m100} shows the relative increase in the MDP, defined as the ratio $\mathrm{MDP}_\sigma/\mathrm{MDP}$ where MDP is given by Eq.~\ref{eq:MDP}. Although the effect is small for low $\sigma_\mu/\mu$ (1\% at $\sigma_\mu/\mu=5\%$), it becomes significant at larger $\sigma_\mu/\mu$ (14\% at $\sigma_\mu/\mu=15\%$) and deteriorates the instrument performance for extreme values (80\% at $\sigma_\mu/\mu=25\%$). The result is independent of the intial MDP.

\section{Conclusions}
The results presented here provide a means of quantifying the errors incurred when using the simple estimators for polarization parameters as well as for their uncertainties. These errors are related to the well-established figure of merit MDP and the reconstructed polarization fraction, making the results easily applicable to any X-ray polarimeter. Unlike some previous works, this analysis does not ignore the correlation between the Stokes parameters $Q$ and $U$.

Additionally, the extent to which the reconstructed fraction $p_\mathrm{r}$ and polarization angle $\psi_\mathrm{r}$ can be used as sufficient statistics is explored. In certain situations, such as for high polarization fraction and high signal-to-background ratio, $Q$ and $U$ are not Gaussian distributed and their likelihood is non-Gaussian, resulting in a significantly different posterior for the polarization parameters when the number of photons is low. In such cases, the Stokes parameters cannot be used at all because $(p_\mathrm{r},\psi_\mathrm{r})$ are not sufficient statistics.

The fact that the reconstructed polarization fraction is biased towards higher values implies that binning a high-statistics data set into smaller subsets to estimate the evolution of the polarization fraction will yield incorrect results when using the simple Gaussian estimator $p_\mathrm{r}$. This work provides a way to decide how coarsely the data must be binned in order to justify the use of Gaussian estimators.

The uncertainties on the estimated parameters are as important as the parameter estimates themselves. When statistics are high $(\mathrm{MDP}/p_\mathrm{r}<0.5)$ the errors are small and can usually be neglected (justifying Gaussian assumptions made when using simple estimators) unless the reconstructed polarization fraction is high, for example, $>0.8$. However, in the statistics-limited regime $(\mathrm{MDP}/p_\mathrm{r}>1)$ the systematic error made from using such simple estimators is non-negligible compared to the statistical uncertainty. Additionally, it is shown how quickly the posterior becomes asymmetric as the statistical power decreases. This effect is strongly dependent on the reconstructed polarization fraction because of the prior $0\leq p_0\leq1$.

Lastly, the effect of uncertainties on the modulation factor $\mu$ is studied. It is shown to be important once the relative uncertainty exceeds $10\%$ and to dominate the performance of an instrument when it is above $20\%$. This is relevant for the optimization of future GRB polarimeters, since they tend to have large uncertainties on $\mu$ due to the difficulty of localizing bursts and measuring the primary GRB spectrum. It also shows that simple Gaussian estimators cannot be used in the high-photon-counts regime for GRB polarimeters when localization uncertainties are high.

\begin{acknowledgements}
This research was supported by the Swedish National Space Board. M. Kiss, M. Pearce, F. Xie and the Referee are thanked for providing constructive feedback on the manuscript.
\end{acknowledgements}

\appendix

\section{Stokes parameters}
\label{appendix_stokes_parameters}

The Stokes parameters can be constructed from
\begin{equation}
\begin{split}
q_i=\cos({2\psi_i})\\
u_i=\sin({2\psi_i})
\end{split}
.\end{equation}
To find an expression for $p_0$ , the means $\mean{q}$ and $\mean{u}$ must be computed.
\begin{equation}
\mean{q}=\int_{0}^{2\pi}\cos(2\psi)f(\psi)\mathrm{d}\psi
\label{eq:1st_moment_q}
,\end{equation}
\begin{equation}
\mean{q}=\frac{\mu_0 p_0S}{2(B+S)}\cos({2\psi_0})
,\end{equation}
where $f(\psi)$ is given by Eq.~\ref{eq:modulation_curve}. Similarly
\begin{equation}
\mean{u}=\frac{\mu_0 p_0S}{2(B+S)}\sin({2\psi_0})
.\end{equation}
The polarization fraction $p_0$ is then given by
\begin{equation}
\begin{split}
p_0=2\sqrt{Q_0^2+U_0^2}/\mu_0\\
\psi_0=\frac{1}{2}\arctan{\frac{U_0}{Q_0}}
\end{split}
,\end{equation}
where $Q_0$ and $U_0$ are the normalized Stokes parameters defined as  
\begin{equation}
\begin{split}
Q_0=\frac{B+S}{S}\mean{q}=\frac{\mu_0 p_0}{2}\cos({2\psi_0})\\
U_0=\frac{B+S}{S}\mean{u}=\frac{\mu_0 p_0}{2}\sin({2\psi_0})
\end{split}
.\end{equation}
The sufficient statistics for data generated from $(Q_0,U_0)$ are $(Q_\mathrm{r},U_\mathrm{r})$ defined as
\begin{equation}
\begin{split}
Q_\mathrm{r}=\frac{1}{S}\sum\limits_{i=1}^Nq_i=\frac{\mu_\mathrm{r} p_\mathrm{r}}{2}\cos({2\psi_\mathrm{r}})\\
U_\mathrm{r}=\frac{1}{S}\sum\limits_{i=1}^Nu_i=\frac{\mu_\mathrm{r} p_\mathrm{r}}{2}\sin({2\psi_\mathrm{r}})
\end{split}
,\end{equation}
where $N=B+S$. In polar coordinates this becomes
\begin{equation}
\begin{split}
p_{\mathrm{r}}=2\sqrt{Q_\mathrm{r}^2+U_\mathrm{r}^2}/\mu_{\mathrm{r}}\\
\psi_{\mathrm{r}}=\frac{1}{2}\arctan{\frac{U_\mathrm{r}}{Q_\mathrm{r}}}
\end{split}
.\end{equation}

\section{Uncertainties on Stokes parameters}
\label{appendix_stokes_uncertainties}

The uncertainties on $Q$ and $U$ as well as their covariance can be derived by computing their second moments.
\begin{equation}
\mean{q^2}=\int_{0}^{2\pi}\cos^2(2\psi)f(\psi)\mathrm{d}\psi=\frac{1}{2}
\label{eq:2nd_moment_q}
.\end{equation}
Similarly
\begin{equation}
\mean{u^2}=\frac{1}{2}
,\end{equation}
and the cross-term is
\begin{equation}
\mean{qu}=\frac{1}{2\pi(S+B)}\int_{0}^{2\pi}\cos(2\psi)\sin(2\psi)\times f(\psi)\mathrm{d}\psi=0
.\end{equation}
Combining Eq.~\ref{eq:1st_moment_q} and Eq.~\ref{eq:2nd_moment_q} yields the standard deviation
\begin{equation}
\sigma_q=\sqrt{\frac{1}{2}-\left(\frac{\mu_0 p_0S\cos({2\psi_0})}{2(B+S)}\right)^2}
.\end{equation}
Here $\sigma_q$ describes the dispersion of $q$ but it is $\sigma_Q$ which is of interest. It is given by
\begin{equation}
\sigma_Q^2=\sigma_q^2\times\frac{N}{S^2}
,\end{equation}
where the division by $S^2$ comes from the definition of normalized Stokes parameters.
\begin{equation}
\sigma_Q=\sqrt{\frac{1}{S}\left(\frac{N}{2S}-\frac{\mu_0^2p_0^2\cos^2(2\psi_0)}{4}\right)}
.\end{equation}
Similarly
\begin{equation}
\sigma_U=\sqrt{\frac{1}{S}\left(\frac{N}{2S}-\frac{\mu_0^2p_0^2\sin^2(2\psi_0)}{4}\right)}
.\end{equation}
It is finally possible to compute the covariance and the correlation coefficient $\rho$
\begin{equation}
\textrm{Cov}(Q,U)=\textrm{Cov}(q,u)\times\frac{N^2}{S^3}=-\frac{\mu_0^2p_0^2}{8S}\sin(4\psi_0)
,\end{equation}
\begin{equation}
\rho=-\frac{S\mu_0^2p_0^2\sin(4\psi_0)}{\sqrt{16N^2-8NS\mu_0^2p_0^2+S^2\mu_0^4p_0^4\sin^2(4\psi_0)}}
.\end{equation}

\section{Relative mean bias}
\label{appendix_relative_mean_bias}

An approximate expression for the relative mean bias can be derived from the Rice distribution
\begin{equation}
f(p_{\mathrm{r}}|p_0)=\frac{p_{\mathrm{r}}}{\sigma^2}\exp{\left(-\frac{p_{\mathrm{r}}^2+p_0^2}{2\sigma^2}\right)}\times I_0\left(\frac{p_{\mathrm{r}}p_0}{\sigma^2}\right)
,\end{equation}
where $\sigma\equiv\sqrt{2N}/\mu_0 S$. The mean $\mean{p_{\mathrm{r}}}$ is given by
\begin{equation}
\begin{split}
\mean{p_{\mathrm{r}}}=\sqrt{\frac{\pi}{2}}\sigma\times\mathrm{exp}{\left(-\frac{p_0^2}{4\sigma^2}\right)} \\
\times \left((1+\frac{p_0^2}{2\sigma^2})I_0\left(\frac{p_0^2}{4\sigma^2}\right)+\frac{p_0^2}{2\sigma^2}I_1\left(\frac{p_0^2}{4\sigma^2}\right)\right)
\end{split}
\label{eq:mean_p_r}
,\end{equation}
where $I_0$ and $I_1$ are modified Bessel functions of the zeroth and first order, respectively. For high-statistics measurements (where $p_0\gg2\sigma$) the Bessel functions can be approximated by expanding them to second order.
\begin{equation}
\lim_{x \to \infty} I_0(x)=\frac{e^x}{\sqrt{2\pi x}}\left(1+\frac{1}{8x}\right)
,\end{equation}
\begin{equation}
\lim_{x \to \infty} I_1(x)=\frac{e^x}{\sqrt{2\pi x}}\left(1-\frac{3}{8x}\right)
.\end{equation}
Finally
\begin{equation}
\lim_{p_0/\sigma \to \infty} \mean{p_{\mathrm{r}}}=\frac{\sigma^2}{2p_0}+p_0
.\end{equation}
So in the limit of high statistics, the relative mean bias $\beta$ is given by
\begin{equation}
\lim_{p_0/\sigma \to \infty} \beta\approx\frac{\sigma^2}{2p_0^2}=\frac{N}{S^2\mu_0^2p_0^2}=\left(\frac{\textrm{MDP}}{4.29p_0}\right)^2
.\end{equation}
By using recursion, $\beta$ can be expressed as a function of $p_\mathrm{r}$ instead of $p_0$ which is not known \textit{a priori}. This simplifies to
\begin{equation}
\begin{split}
\beta\approx\left(\frac{\textrm{MDP}}{4.29p_0}\right)^2=\sum_{n=1}^\infty C_nx^{2n}\\
=\sum_{n=1}^\infty\frac{1}{n+1}{2n \choose n}x^{2n}=\frac{1-2x^2-\sqrt{1-4x^2}}{2x^2}
\end{split}
,\end{equation}
where $x=\textrm{MDP}/4.29p_{\mathrm{r}}$ and $C_n$ is the $n$th Catalan number.

\section{Gaussian uncertainties on polarization parameters}
\label{appendix_uncertainties}
The uncertainty on $p_\mathrm{r}$ can be derived by using the standard uncertainty propagation formula
\begin{equation}
\sigma_{p_\mathrm{r}}^2=\bigg|\frac{\mathrm{d}p_\mathrm{r}}{\mathrm{d}Q_\mathrm{r}}\bigg|^2\sigma_Q^2+\bigg|\frac{\mathrm{d}p_\mathrm{r}}{\mathrm{d}U_\mathrm{r}}\bigg|^2\sigma_U^2+2\bigg|\frac{\mathrm{d}p_\mathrm{r}}{\mathrm{d}Q_\mathrm{r}}\bigg|\bigg|\frac{\mathrm{d}p_\mathrm{r}}{\mathrm{d}U_\mathrm{r}}\bigg|\mathrm{Cov}(Q,U)
,\end{equation}
yielding
\begin{equation}
\sigma_{p_\mathrm{r}}=\frac{2}{\mu_r}\sqrt{\frac{1}{S}\left(\frac{N}{2S}-\frac{\mu_0^2p_0^2}{4}\right)}
,\end{equation}
and similarly for $\psi_\mathrm{r}$
\begin{equation}
\sigma_{\psi_{\mathrm{r}}}=\frac{1}{\sqrt{2}\mu_\mathrm{r} p_{\mathrm{r}}}\frac{\sqrt{N}}{S}
.\end{equation}

\end{document}